\documentclass[conference]{IEEEtran}

\IEEEoverridecommandlockouts

\usepackage{cite} 
\usepackage{amsmath,amssymb,amsfonts}
\usepackage{algorithmic}
\usepackage{graphicx}
\usepackage{textcomp}
\usepackage{booktabs}
\usepackage{pifont}      % For checkmarks and xmarks
\usepackage{xcolor}      % For color control
\usepackage{colortbl} 

\usepackage{listings}
\usepackage{xcolor}

\definecolor{badhl}{RGB}{255,235,235}
\definecolor{codebg}{RGB}{245,245,245}
\definecolor{codeframe}{RGB}{150,150,150}
\definecolor{codekw}{RGB}{0,128,0}
\definecolor{codecom}{RGB}{70,120,170}
\definecolor{codestr}{RGB}{160,80,20}
\definecolor{codelineno}{RGB}{90,90,90}

\lstdefinestyle{mypython}{
    language=Python,
    backgroundcolor=\color{codebg},
    basicstyle=\ttfamily\fontsize{6}{7}\selectfont,
    keywordstyle=\color{codekw}\bfseries,
    commentstyle=\color{codecom}\itshape,
    stringstyle=\color{codestr},
    identifierstyle=\color{black},
    numbers=left,
    numberstyle=\ttfamily\fontsize{5.5}{6}\selectfont\color{codelineno},
    stepnumber=1,
    numbersep=6pt,
    showstringspaces=false,
    showspaces=false,
    showtabs=false,
    keepspaces=true,
    columns=fullflexible,
    breaklines=true,
    breakatwhitespace=false,
    tabsize=4,
    xleftmargin=6pt,
    xrightmargin=3pt,
    aboveskip=2pt,
    belowskip=2pt,
    frame=single,
    rulecolor=\color{codeframe},
    framesep=2pt,
    framerule=0.3pt
}

\usepackage{pifont}
\usepackage{booktabs}
\usepackage{makecell}
\usepackage[table]{xcolor}
\newcommand{\cmark}{\textcolor{green!60!black}{\ding{51}}} 
\newcommand{\xmark}{\textcolor{red!70!black}{\ding{55}}} 
\definecolor{codebg}{RGB}{248,248,248}
\definecolor{badhl}{HTML}{FDE2E2} % soft red for error line
\definecolor{goodhl}{HTML}{E8F5E9} % soft green for fixed 
\definecolor{codebg}{gray}{0.97}    
\definecolor{LightGray}{gray}{0.93}
\usepackage{array}
\def\BibTeX{{\rm B\kern-.05em{\sc i\kern-.025em b}\kern-.08em
    T\kern-.1667em\lower.7ex\hbox{E}\kern-.125emX}}

\usepackage{listings}
\usepackage{xcolor}
\usepackage{makecell}

\begin{document}

\title{Customized User Plane Processing via Code Generating AI Agents for Next Generation Mobile Networks}

\author{
    \IEEEauthorblockN{Xiaowen Ma\IEEEauthorrefmark{1}, Onur Ayan\IEEEauthorrefmark{1}, Yunpu Ma\IEEEauthorrefmark{2}, Xueli An\IEEEauthorrefmark{1}}
    \IEEEauthorblockA{\IEEEauthorrefmark{1}Huawei Technologies Duesseldorf GmbH
    \\\{xiaowen.ma, onur.ayan, xueli.an\}@huawei.com}
    \IEEEauthorblockA{\IEEEauthorrefmark{2}Ludwig-Maximilians-University Munich
    \\\{yunpu.ma\}@ifi.lmu.de}
}

\maketitle

\begin{abstract}

Generative AI is envisioned to have a crucial impact on next generation mobile networking, making the sixth generation (6G) system considerably more autonomous, flexible, and adaptive than its predecessors. By leveraging their natural language processing and code generation capabilities, AI agents enable novel interactions and services between networks and vertical applications. A particularly promising and interesting use case is the customization of connectivity services for vertical applications by generating new customized processing blocks based on text-based service requests. More specifically, AI agents are able to generate code for a new function block that handles user plane traffic, allowing it to inspect and decode a protocol data unit (PDU) and perform specified actions as requested by the application. In this study, we investigate the code generation problem for generating such customized processing blocks on-demand. We evaluate various factors affecting the accuracy of the code generation process in this context, including model selection, prompt design, and the provision of a code template for the agent to utilize. Our findings indicate that AI agents are capable of generating such blocks with the desired behavior on-demand under suitable conditions. We believe that exploring the code generation for network-specific tasks is a very interesting problem for 6G and beyond, enabling networks to achieve a new level of customization by generating new capabilities on-demand.
\end{abstract}

\begin{IEEEkeywords}
Agentic AI, Communication Protocols, Network Automation, Code Generation
\end{IEEEkeywords}

\begin{figure*}[!t]
\centering
\includegraphics[width=2.0\columnwidth]{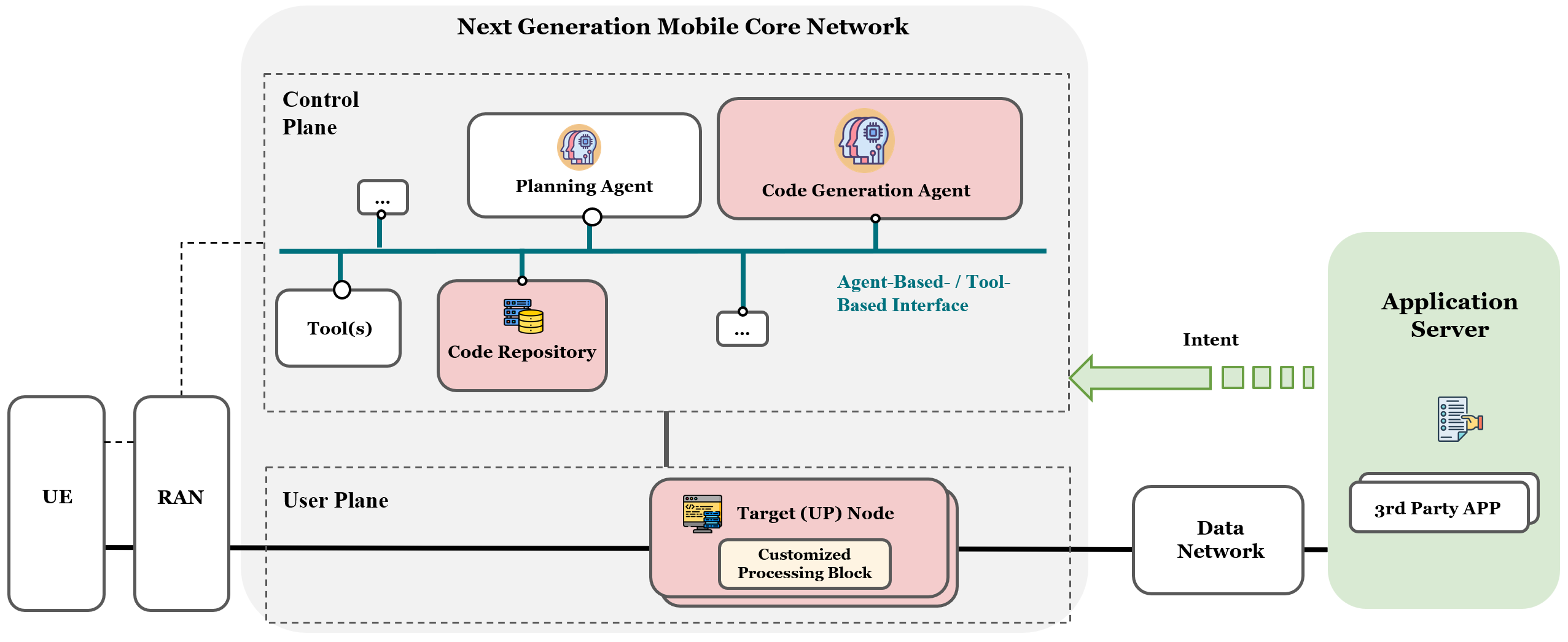}
\caption{System architecture of the proposed AI-agent-enabled framework for on-demand Customized Processing Block generation within the mobile core network.}
\label{fig:system_overview}
\end{figure*}

\section{Introduction and Related Work}

Recent advances in artificial intelligence (AI) and machine learning (ML) have significantly impacted various sectors including robotics, healthcare, biotechnology, transportation, and entertainment. Generative AI, which is one of the most popular sub-categories of AI research, has gained popularity, particularly thanks to chatbots for question and answering tasks, and code generation for software development, empowered by large language models (LLMs). AI agents turned LLMs into software components that can interact with external tools, using various APIs, e.g., for searching web, querying a database, making a hotel reservation, or placing a food order. 

As the scale of agentic AI systems grew, academia and industry have identified a set of challenges and open research questions around interoperability. This created a need for communication protocols that are specifically designed for agent-to-agent communication and agent-to-x communication, where the ``x'' can be a database, a server, or any other component that does not necessarily have natural language processing capabilities. To address these challenges, several protocols have been proposed, including but not limited to Agent2Agent (A2A), agent communication protocol (ACP), and model context protocol (MCP)\footnote{One can refer to \cite{ietfaiagent, etsieni056} for further information about agent communication protocols.}.

Agentic AI technology is expected to transform the telecommunications sector significantly, not only due to the increased demand for mobile data traffic, e.g., for user-chatbot interaction for simple question and answering. On the contrary, AI agents are envisioned to be natively integrated into the next generation mobile communication systems. 

Despite being a relatively new research area, the existing literature already contains several works focusing on the use cases, trends, and challenges of utilizing generative AI in telecommunications \cite{boateng2025survey, guo2025large, yang2025decision, qayyum2025llm}. One of the existing prominent papers is \cite{maatouk2025teleqna}, which conducts a study where they generate an evaluation dataset based on various specifications, research papers and other data sources. In their work, authors conclude that LLMs are able to compete with active professionals when it comes to answering telecom-specific questions. 

Moreover, \cite{zou2025telecomgpt} focus on the training of domain-specific LLMs and discuss its main challenges. A very interesting and promising task they identify as relevant for generative AI-empowered networks is code generation. In their work, authors challenge LLMs for generating code that are taken from open-source files with relevance to telecom domain. They prompt LLMs to generate a function (e.g., in Python language) or perform code completion (i.e., code infilling), where the original file is considered as ground truth and the semantic similarity between the generated and original code is used to quantify the performance. However, function generation can be considered as a very isolated sub-problem, decoupled from a particular use case when it comes to applications and requirements in next generation mobile networking. In addition, although employing the code similarity w.r.t. a ground truth is insightful, it may not always capture the potential errors introduced by LLMs for code generation, which has been categorized in \cite{song2023empirical}.

In this work, we focus on LLM-based code generation for user plane processing. More specifically, we focus on the interaction between a third party application and the mobile network, where the network is provided with a customized protocol description in human-readable text. We prompt AI agents to generate a processing entity with corresponding interfaces to receive packets and perform various actions, as requested. In such a setting, we conduct a case study based on three custom protocol descriptions in different complexity levels and evaluate the code generation performance w.r.t. protocol-specific functionality. This includes sending actual packets between transmitters and receivers, where the system logs are analyzed to capture any erroneous behavior. We analyze the functional correctness of the generated code under four different settings, while we vary the selected model, prompt content, and input code templates. To the best of our knowledge, this is the first work that evaluates the code generation performance of LLMs in networking domain, where the evaluation is not based on metrics quantifying similarity between a reference code and output code, but based on the functional correctness of the generated source code in a realistic deployment scenario.

\section{Customized Processing Block Generation}
\label{sec:scenario_and_generation}

\subsection{System Architecture}
\label{subsec:system_arch}
We consider a mobile core network that comprises LLM-empowered AI agents\footnote{Integrating AI agents into mobile core network has been proposed in the existing literature, such as \cite{etsieni051, tong2025acore, tothfalusi2025taco}.}. Similar to 5G network architecture, where the over-the-top (OTT) application can interact with the mobile communication (via the application function (AF)), we assume an interface between the network and the application, through which 3\textsuperscript{rd} party applications can provide service requests to the mobile system. Different than in 5G, we assume that this request can contain parts that are encoded in human-readable text format.

As depicted in Figure~\ref{fig:system_overview}, the mobile core network contains three key components colored in red, namely, a code generation agent (CGA), a code repository (CR), and user plane (UP) node(s). Together with these, the figure also depicts other entities such as further network AI agents (e.g., a planning agent) with different roles and responsibilities, as well as tools that can be used by network AI agents to perform various network control tasks, such as session management, mobility management etc.\footnote{The specifics of entities from Figure \ref{fig:system_overview} that are colored in white as well as the interface design between them are out of this work's scope.}.

\textbf{CGA} is responsible for generating an executable code upon the reception of a service request sent by the OTT application. In other words, the CGA is requested to generate a \textit{customized processing block (CPB)}, the behavior of which is characterized by a \textit{protocol description} contained in the service request. In this work, we limit the scope to tasks that are related to user plane packet processing, e.g., for an IP flow between an application server and a user equipment (UE). More importantly, we assume that the requested behavior does not belong to already supported capabilities of the mobile network, which further implies that the network must generate these in an on-demand fashion in order to provide the requested service. 

\textbf{CR} is a centralized code knowledge base that stores network-related code samples. Examples of such would be various common code snippets, modules, or examples that are useful for in-network processing. The CR serves as a shared code database for other entities in a mobile communication system, such as the code generation agent, which may utilize these for model training or in-context learning. An example realization of the CR in a practical setup is to implement it as an MCP server exposing a list of \textit{resources}\footnote{The MCP specification defines ``resources" as passive data sources that provide read-only access to information for context, such as file contents, database schemas, or API documentation.\cite{mcpwebsite}}.

\textbf{Target (UP) node} is a network node on the data path between application server and UE, through which the data packets travel between source and destination. In our architecture, the target UP node plays a central role, since it is the location where the CPB generated by the code generation agent is deployed\footnote{5G contains user plane function (UPF) in core network architecture to processing user plane packets. We do not necessarily assume that target node is UPF.}.

\subsection{Protocol Description and Prompt Design}
\label{subsec:protocol_description}

Communication protocols enable communication between devices across diverse types of networks and incorporate mechanisms for error detection, congestion control, packet routing, and security. They are essential for information exchange between devices and components, as they define the set of rules and conventions, ensuring interoperability and reliable information exchange. 
There are various mechanisms proposed and/or supported for mobile communications that make use of standardized protocols for in-network processing and decision-making. Especially for providing improved service experience and system efficiency for vertical applications, a vast amount of mechanisms and algorithms have been proposed aiming to improve the service experience and system efficiency for vertical applications. Some examples for extended reality and media (XRM) services can be found in \cite{3gppTR23700}. 

According to \cite{kurose2021computer}, a protocol defines the \textit{format} (i.e., syntax and semantics) and the \textit{order} of messages exchanged between two or more communicating entities (i.e., timing), as well as the \textit{actions} taken on the transmission and/or receipt of a message or other event. This implies that the prompt to the code generation agent contains a combination of these elements. For instance, the ``Simple Transmission Protocol'', which is a custom protocol that we defined for our case study later in Section \ref{sec:protocol_cases}, defines a custom packet type with four fields and describes their length and location. This information is particularly important for the parts of CPB code that extract the information, when a new packet is received.

As described by the authors in \cite{kurose2021computer}, a protocol defines actions taken on the transmission and/or receipt of messages. Typical examples are \texttt{HTTP PUT} method where the recipient creates a new resource or replaces a representation of the target resource with the request content, returning a \texttt{201 (Created)} status code after successfully creating a resource \cite{rfc7231}, and the \texttt{Send-Terminate-Ack} action of the \textit{Point-to-Point Protocol (PPP)} that triggers the transmission of the \texttt{Terminate-Request} packet, which requests its receiver to terminate the connection\cite{rfc1171}.

Having said that, documents similar to \cite{rfc7231} and \cite{rfc1171} typically describe a high level behavior, without providing any implementation specific details. However, as we are interested in generating code that implements an executable CPB, our prompt contains implementation-specific actions that characterize internal processing logic of the CPB. To name an example, the prompt can contain functional requirements such as requesting the generated processing entity to maintain a transmission queue with first-come, first-serve (FCFS) strategy. Moreover, it can describe a policy that determines when an incoming data packet is admitted into the transmission queue or when a packet is discarded. In addition, we include examples in the prompt that facilitate in-context learning, e.g., semantic information for information elements in a data packet and example state-action pairs describing the correct actions to be taken by the CPB. 

To improve the code generation accuracy, we include another important component alongside the prompt, which we refer to as \textit{baseline code}. This element contains one or multiple code snippets that are relevant for the code generation task and is located at the CR to be fetched by the code generation agent. This leverages the approach proposed in \cite{li2023motcoder}, in which the prompt contains the names and doc-strings that are necessary to fulfill the code generation task. As discussed later in Section \ref{sec:exp_results}, including examples in the protocol description as well as baseline code in the prompt have significant impact on the correctness of the output code.

\subsection{Customized Processing Block Generation and Execution}
\label{subsec:cpb_generation}

\begin{figure}[!t]
\centering
\includegraphics[width=1.0\columnwidth]{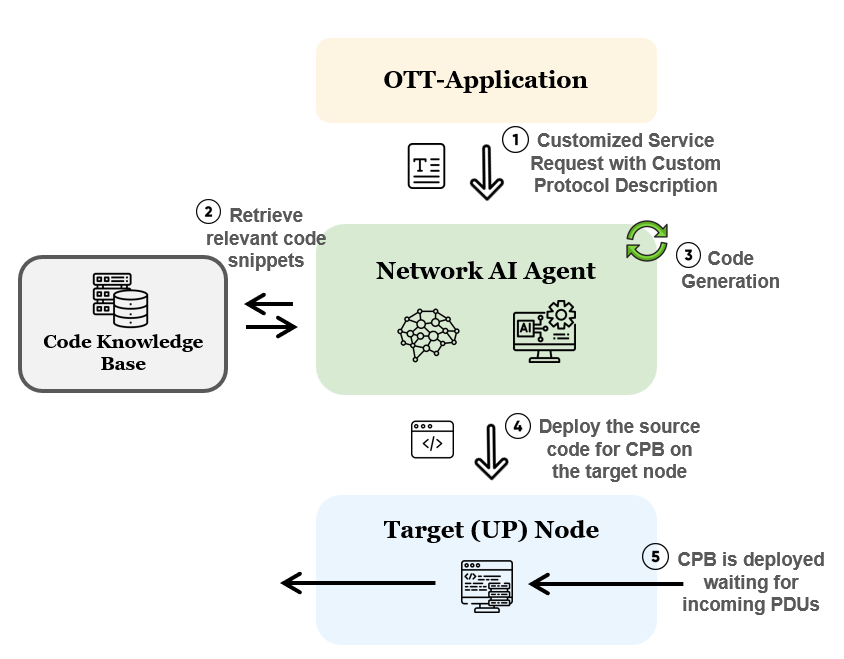}
\caption{Processing flow of the proposed on-demand Customized Processing Block (CPB) generation. Upon receiving a customized service request with a protocol description from the OTT application (\ding{172}), the Network AI Agent retrieves relevant code snippets from the Code Knowledge Base (\ding{173}), generates the source code for the CPB (\ding{174}), and deploys it on the target user-plane node (\ding{175}). The deployed CPB then waits for incoming PDUs and executes the requested processing function (\ding{176}).}

\label{fig:use_case}
\end{figure}
In our framework, the generation and execution of the CPB represent the core stages, in which protocol descriptions are transformed into executable in-network processing entities. The code generation agent possesses the flexibility to select, combine, and invoke tools within the available toolset, enabling it to autonomously generate a customized processing block that implements the protocol description. The code generation process integrates the protocol description and baseline code into a single input prompt, where the description defines the target behavior and the baseline provides reusable socket and communication logic. The output is a fully functional Python-based customized processing block that is automatically executed within the experimental environment.

The processing flow of the proposed on-demand Customized Processing Block (CPB) generation is illustrated in Figure~\ref{fig:use_case}. It highlights the end-to-end interaction among the OTT application, Network AI Agent, code knowledge base, and the target user-plane node, showing how the proposed framework can automatically translate a customized service request into an executable and adaptive processing block, effectively bridging high-level network specifications with real-time operational code in mobile communication systems.

\section{Case Study and Main Results}
\label{sec:exp}

\subsection{Selected Use Cases}
\label{sec:protocol_cases}
We conduct a case study to evaluate the ability of LLM-empowered AI agents to generate networking-specific source code. To that end, we consider three custom protocol descriptions that represent common packet processing patterns found in computer networking:

\noindent\textbf{Simple Transmission Protocol (STP):}
The code generation agent is prompted to generate a customized processing block that maintains a first-come-first-served (FCFS) packet queue, where packets contain a \textit{priority} field. The STP description requires the CPB to admit an incoming data packet to the transmission (TX) queue based on a fixed priority threshold before sending it to the destination or discard otherwise.

\noindent\textbf{Congestion-Control Protocol (CC):}
This protocol introduces a new type of packet, namely a \textit{control packet} sent by a network controller, which the CPB should be able to decode and parse. The purpose of the control packet is to inform the CPB about the dynamic network congestion status, i.e., a binary indicator whether congestion exists or not. In addition, the CPB receives data packets to be forwarded to the destination, which carry a priority field as in STP. In contrast to STP, the CPB is requested to consider the packet priority only if the network is congested. More specifically, a packet below the priority threshold is admitted to the TX queue only if the network's congestion status is false (not-congested).

\noindent\textbf{Publish-Subscribe Protocol (Pub-Sub):}
The protocol description describes a custom publish-subscribe protocol where the code generation agent is prompted to output the source code implementing a pub-sub message broker. The broker receives three types of packets, namely, a \texttt{SUBSCRIBE} packet indicating a request to subscribe to a specific topic, an \texttt{UNSUBSCRIBE} packet to remove a subscription, and a \texttt{PUBLISH} packet which contains published information on a specific topic. In addition, the broker is required to acknowledge a (un-)subscription with an \textit{ACK} to the sender.

\subsection{Experimental Setting}
\label{sec:exp_setup}

\noindent\textbf{Scenario configurations:}
We consider four scenarios for each protocol use case to isolate the impact of model selection, the availability of a baseline code at the code generation agent to utilize as a starting point, and the provision of examples within the prompt for in-context learning. This is summarized in Table~\ref{tab:scenarios}, where Scenario~1, i.e., S1, uses \texttt{Gemini-2.5-flash} with the provision of baseline code as well as examples in the prompt. Scenario~2 (S2) removes the baseline code only, in contrast to S1, to study the importance of providing a code skeleton to the CGA to support the generation\footnote{Findings in \cite{li2023motcoder} suggest that the provision of relevant code snippets or skeleton is expected to improve the accuracy of the output code.}. Scenario~3 (S3) re-introduces the baseline code, while removing the examples from the prompt to isolate the role of examples in the prompt. Scenario~4 (S4) has both the baseline code and examples, however, employs \texttt{Gemini-2.0-flash} instead of \texttt{Gemini-2.5-flash}. Since switching to \texttt{Gemini-2.0-flash} already resulted in noticeably lower performance, we chose not to extend further scenarios in which the supporting elements for code generation (i.e., baseline code or examples) are removed.

\begin{table}[htbp]
\renewcommand{\arraystretch}{1.25}
\caption{Scenario configuration and Feature Comparison}
\centering
\setlength{\tabcolsep}{5pt}
\begin{tabular}{>{\columncolor{gray!10}}lccc}
\rowcolor{gray!20}
\textbf{Scenario} & \textbf{Model}  & 
\textbf{\makecell[c]{Baseline Code\\Provided}}  & \textbf{\makecell[c]{Examples in \\ Protocol Description}} \\
\toprule
S1   & Gemini-2.5-flash & \cmark & \cmark \\[2pt]
\rowcolor{gray!05}
S2  & Gemini-2.5-flash & \xmark & \cmark \\[2pt]
S3  & Gemini-2.5-flash & \cmark & \xmark \\[2pt]
\rowcolor{gray!05}
S4  & Gemini-2.0-flash & \cmark & \cmark \\[2pt]
\bottomrule
\end{tabular}
\label{tab:scenarios}
\end{table}

\noindent\textbf{Functional validation:}
\cite{hou2024large} reviews a large amount prior works that study LLM-based code generation. As their study reveal, a common approach to quantify code generation performance is to employ a similarity metric, which compares the output code to a reference (benchmark) code. Although this is sufficient and well-suited for generic code generation tasks, our goal is to generate a processing entity that should operate correctly. To that end, instead of evaluating based on semantic similarity, we adopt a different approach that is based on processing of actual packets, summarized as follows. 

After the generation of the source code, we execute it on a physical host, to which transmitter(s) are sending actual data packets via a TCP connection. We log use-case-specific events, such as the reception, transmission, or discard of packets together with other relevant information, such as timestamps or packet content. Based on the logged information, we apply a custom validation function to check whether (i) the generated code is executed successfully, (ii) binds to the designated TCP endpoints, (iii) exchanges packets that match the specified formats (e.g., control type, payload encoding), and (iv) satisfies the expected protocol logic. For example, in the Pub-Sub protocol, the broker must update the subscription state and forward the published information on a specific topic to all of its subscribers.

\begin{table}[htbp]
\renewcommand{\arraystretch}{1.25}
\caption{Error types}
\centering
\setlength{\tabcolsep}{4pt}
\begin{tabular}{>{\columncolor{gray!10}}p{4.2cm}!{\color{gray!40}\vrule width 0.8pt}p{3.8cm}}
\rowcolor{gray!20}
\multicolumn{2}{c}{\textbf{Error Types}} \\
\toprule
Condition Error (CE) & \makecell[l]{Missing condition\\Incorrect condition} \\[2pt]
\rowcolor{gray!05}
Constant Value Error (CVE) & \makecell[l]{Constant value error} \\[2pt]
Reference Error (RE) & \makecell[l]{Wrong method/variable\\Undefined name} \\[2pt]
\rowcolor{gray!05}
Code Block Error (CBE) & \makecell[l]{Incorrect code block\\ Missing code block} \\
Incomplete Code / Missing Statements (IC/MS) & \makecell[l]{Missing one statement\\Missing multiple statements} \\[2pt]
\rowcolor{gray!05}
Method Call Error (MCE) & \makecell[l]{Incorrect function arguments\\Incorrect method call target} \\
Operation/Calculation Error (O/CE) & \makecell[l]{Incorrect arithmetic operation\\Incorrect comparison operation} \\
\bottomrule
\end{tabular}
\label{tab:errortype}
\end{table}

\noindent\textbf{Classification of code generation errors:}
Failures are classified using a consolidated taxonomy defined in \cite{song2023empirical}, based on which, the errors in our results fall under seven types: Condition Error (CE), Constant Value Error (CVE), Reference Error (RE), Incomplete Code/Missing Statements (IC/MS), Method Call Error (MCE), Operation/Calculation Error (O/CE) and Code Block Error (CBE). Each of these categories is defined in Table~\ref{tab:errortype}, which summarizes the representative error type and its sub-type.

As a concrete instance, Figure~\ref{fig:pubsub_re_example} shows an error example of \emph{``Operation/Calculation Error - Incorrect Arithmetic Operation (IAO)''} in the Pub–Sub  implementation: the \texttt{PUBLISH} packet fields are serialized in the wrong order, with \emph{payload\_len} placed before the topic.

\begin{figure}[t]
\centering
\begin{minipage}{0.98\columnwidth}
\begin{lstlisting}[style=mypython]
def _send_publish_message(self, ...):
    """Sends a PUBLISH packet to a subscriber."""
    ...
    # Correct order should be:
    # Control Type (1) + Topic Len (2) + Topic Name
    # (N) + Payload Len (4) + Payload (M)
    header = struct.pack("!BH I", CONTROL_TYPE_PUBLISH, topic_len, payload_len)
    # <-- wrong: payload_len before topic
    packet = header + topic_name_bytes + payload_bytes
    subscriber_sock.sendall(packet)
\end{lstlisting}
\end{minipage}
\caption{Example of the error type \emph{``Operation/Calculation Error -- Incorrect Arithmetic Operation (IAO)''} observed in a Pub--Sub protocol use case.}
\label{fig:pubsub_re_example}
\end{figure}

\vspace*{-3mm}

\subsection{Results}
\label{sec:exp_results}

We repeat each scenario defined in Table~\ref{tab:scenarios} 20 times per selected protocol, i.e., STP, CC, and Pub-Sub. We consider a binary classification of \texttt{PASS} and \texttt{FAIL} based on custom scripts that validates the logged information and actions taken by the CPB and/or received packets by other entities. Subsequently, we compare the actions taken during our experiments with the correct actions (e.g., whether all subscribers received a published packet on a topic they subscribed to). Only if there is a complete overlap, then we classify the run as \texttt{PASS}. Otherwise, it is classified as \texttt{FAIL} indicating incorrect behavior.

\begin{figure*}[!t]
\centering
\includegraphics[width=0.33\textwidth]{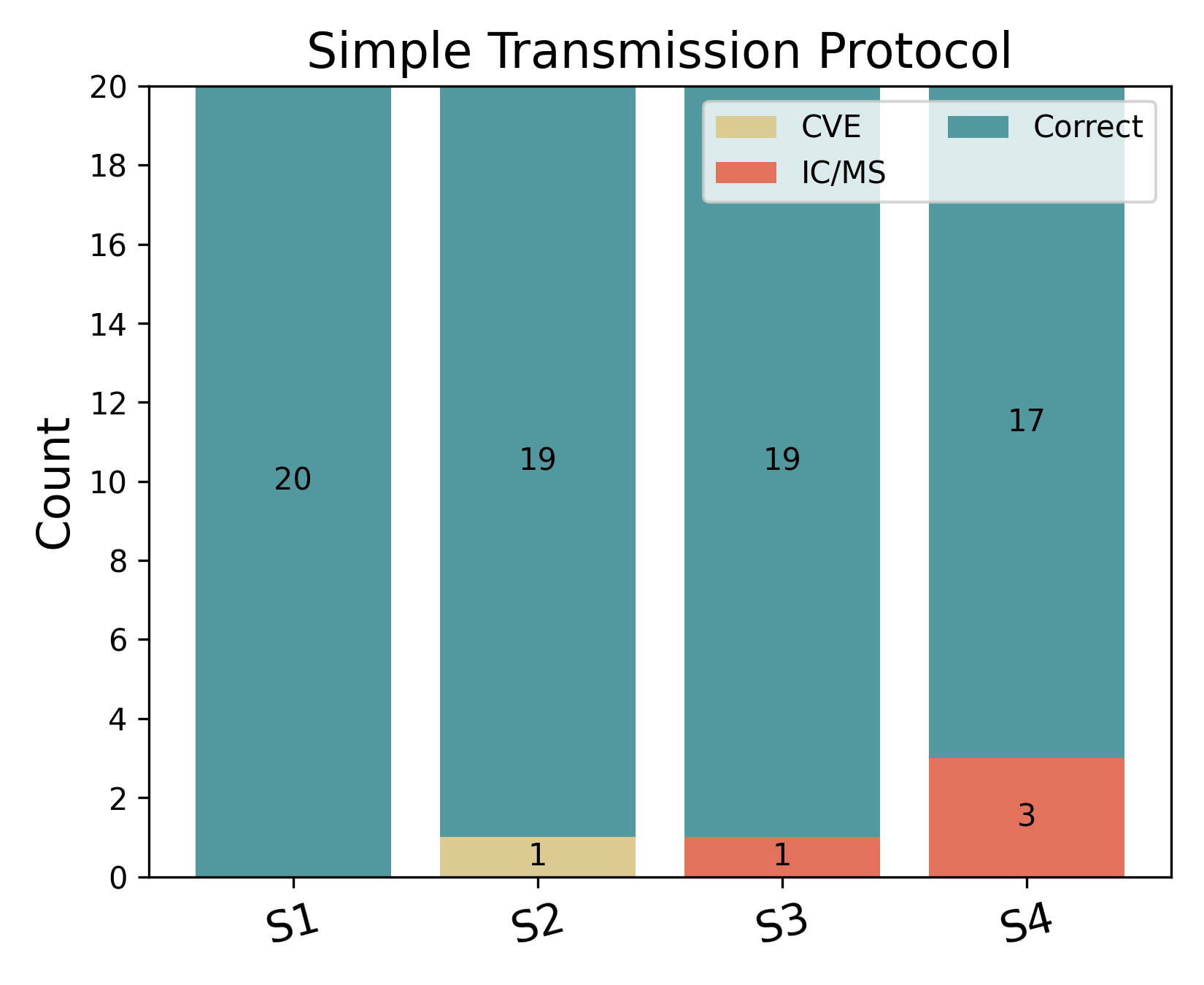}\hfill
\includegraphics[width=0.33\textwidth]{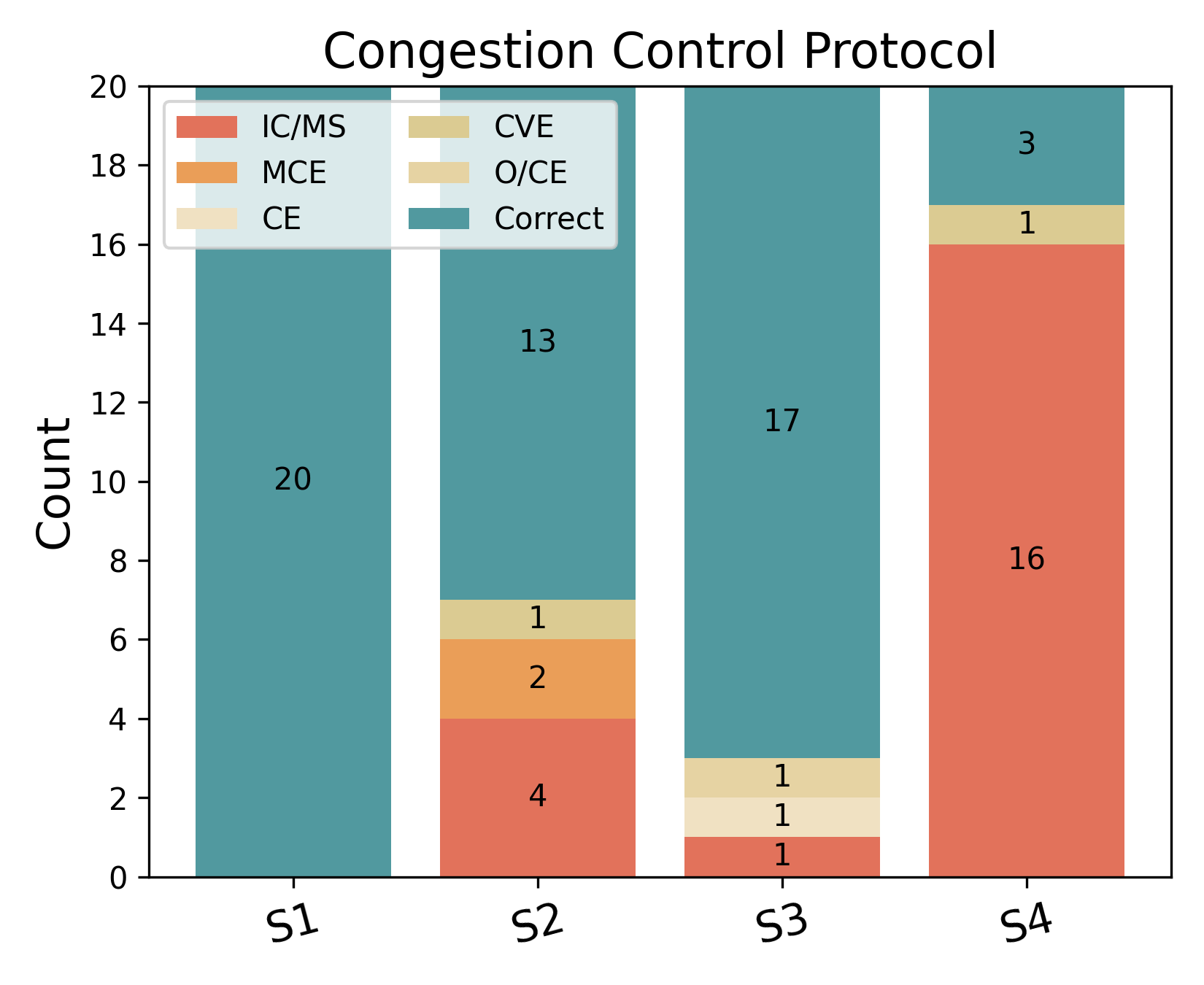}\hfill
\includegraphics[width=0.33\textwidth]{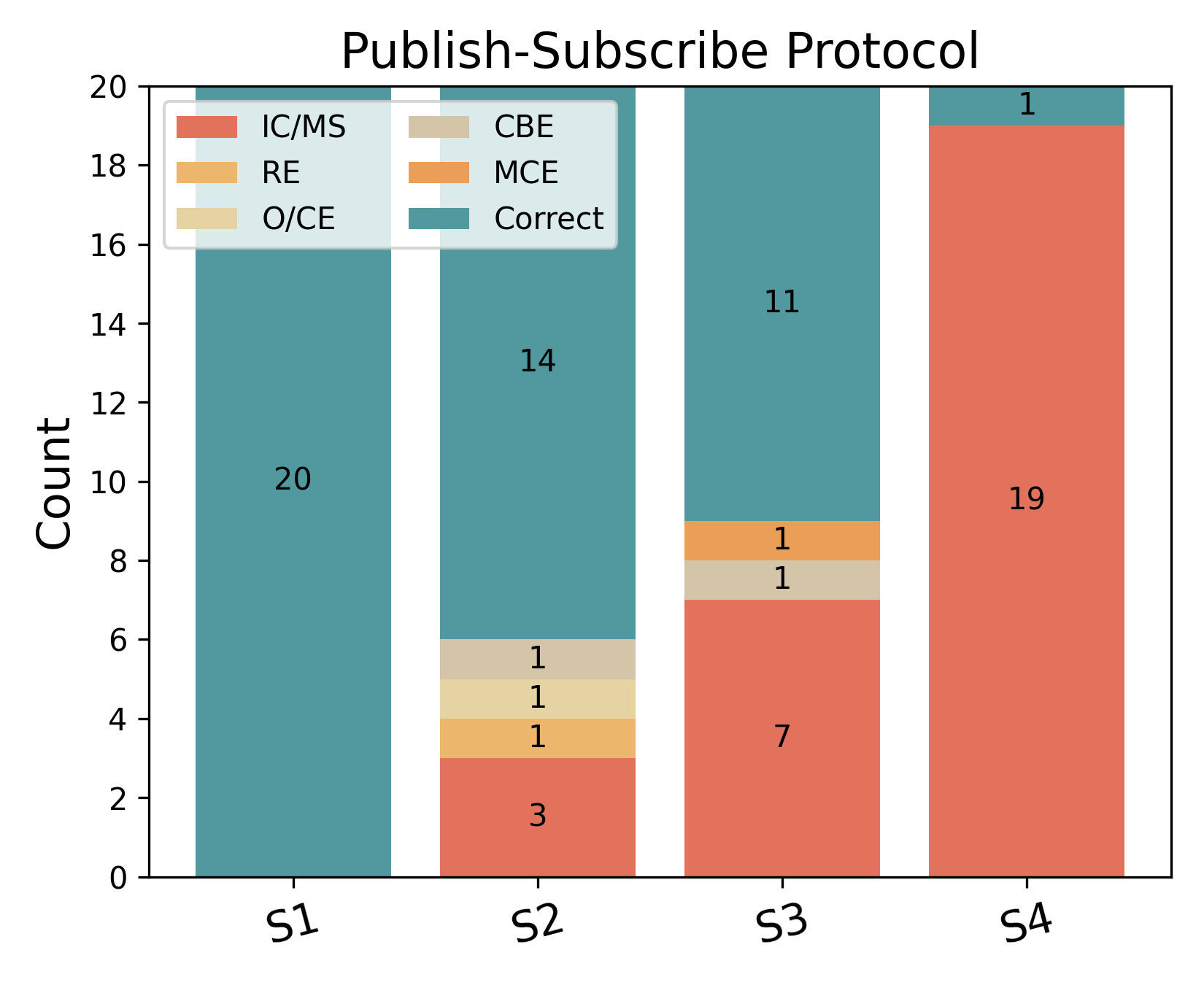}
\caption{Error composition across three protocols under different configurations.}
\label{fig:error}
\end{figure*}

Figure~\ref{fig:error} shows a detailed breakdown of errors that primarily lead to erroneous behavior. Across the three selected protocols of our case study, only the configuration in Scenario~1 yields correct behavior (0/20 failures or equivalently 20/20 passes). Removing the baseline code (S2) or removing the examples in the protocol description (S3) degrades performance, most notably as the complexity of the protocol description increases. Moreover, the results reveal that using a less advanced model, i.e., \texttt{Gemini-2.0-flash}, which corresponds to S4, has a significant impact as well. This shows the difficulty in handling multi-step logic and reliably executing complex protocols. 

For the STP, errors remain limited: without baseline code, only a CVE occurs, while the absence of examples in the protocol description results in a single IC/MS-Missing one statement; with the earlier model, additional errors emerge, including multiple IC/MS-Missing one statement and one IC/MS-Missing multiple statements. These localized and lightweight errors indicate that the simple task can be largely solved even without baseline code or example provision, consistent with its minimal state requirements and straightforward packet handling. 

For the CC, the absence of baseline code leads to a combination of IC/MS-Missing one statement, MCE-Incorrect method call target and CVE. When examples are omitted, the error distribution primarily involves IC/MS-Missing one statement, CE-Missing condition and O/CE-Incorrect arithmetic operation, suggesting that examples play an important role in capturing call sequences and state transitions. Under the weaker model, failures are widespread and mainly of IC/MS-Missing multiple statements and CVE. 

For the Pub-Sub protocol, the absence of baseline code primarily results in IC/MS-Missing one statement, RE-Wrong method/variable, O/CE-Incorrect arithmetic operation and CBE-Incorrect code block. When examples are omitted, the error distribution primarily involves IC/MS-Missing one statement, and occasionally includes MCE-Incorrect method call target and CBE-Incorrect code block. In contrast, the original setting produces no errors, whereas the weaker model exhibits widespread failures, with all errors attributable to IC/MS-Missing multiple statements.

The results indicate that different elements in the code generation input prompt play distinct roles. Baseline code provides structural support: it constrains the behavior of generated code, reducing incomplete implementations and garbage code. Examples in the description provide procedural support: they guide the model in producing correct call sequences, especially for admission control and queue handling, thereby mitigating MCE and residual IC/MS. Model capacity determines whether these elements can be combined into coherent multi-step behavior. Newer models show stronger reasoning and code generation abilities, enabling them to maintain protocol state, follow packet formats, and achieve correct end-to-end semantics more reliably. Finally, task complexity influences sensitivity: simpler protocols remain robust even when one support axis (baseline code or example prompt) is weakened, while more complex protocols rely on both.

\section{Conclusion and Future Work}
LLM-based code generation refers to the process of using an LLM to generate programming code based on natural language inputs or other programming-related prompts. Code generation is a promising use case for LLM-empowered network AI agents, as it allows them to generate new capabilities in an on-demand fashion. In this work, we employed LLMs to generate in-network processing blocks to apply customized handling of data packets. Having a scenario in mind, in which the vertical application describes a custom user plane behavior in natural language, we employed LLMs that are commercially available to generate source code that implements the described logic. We validated the correctness of the generated code by inspecting the actions taken during our real-life experiments based on actual transmission of data packets. Our results suggest that if the model selection and prompt design are performed carefully, LLM-based code generation offers a promising step toward more autonomous and adaptive next generation mobile networks. 

As a part of future work, we plan to extend our study to include further models that are fine-tuned for code generation tasks, such as Code Llama, Qwen3-Coder. Moreover, we plan to create a training dataset with network-specific programming tasks to fine-tune the selected subset of these models to investigate the importance of fine-tuning for telecom-specific code generation tasks. We believe that code generation capability of AI agents will unlock many novel and innovative use cases for mobile networking and play a central role for designing fully autonomous and self-optimizing next generation mobile networks.

\bibliographystyle{IEEEtran}  
\bibliography{IEEEabrv,refs.bib}  

\vspace{12pt}

\end{document}